# Gender bias in academic recruitment[1]

Giovanni Abramo (corresponding author)
*Laboratory for Studies of Research and Technology Transfer*
*Institute for System Analysis and Computer Science (IASI-CNR)*
*National Research Council of Italy*
ADDRESS: Consiglio Nazionale delle Ricerche
Istituto di Analisi dei Sistemi e Informatica
Via dei Taurini 19, 00185 Roma – ITALY
tel. and fax +39 06 72597362, giovanni.abramo@uniroma2.it

Ciriaco Andrea D'Angelo
*Department of Engineering and Management-University of Rome "Tor Vergata"*
ADDRESS: Dipartimento di Ingegneria dell'Impresa
Università degli Studi di Roma "Tor Vergata",
Via del Politecnico 1, 00133 Roma – ITALY
tel. and fax +39 06 72597362, dangelo@dii.uniroma2.it

Francesco Rosati
*Department of Management Engineering-Technical University of Denmark*
ADDRESS: Technical University of Denmark
Produktionstorvet Building 426
2800 Kgs. Lyngby - Denmark
tel +45 45256021, frro@dtu.dk

## Abstract

It is well known that women are underrepresented in the academic systems of many countries. Gender discrimination is one of the factors that could contribute to this phenomenon. This study considers a recent national academic recruitment campaign in Italy, examining whether women are subject to more or less bias than men. The findings show that no gender-related differences occur among the candidates who benefit from positive bias, while among those candidates affected by negative bias, the incidence of women is lower than that of men. Among the factors that determine success in a competition for an academic position, the number of the applicant's career years in the same university as the committee members assumes greater weight for male candidates than for females. Being of the same gender as the committee president is also a factor that assumes greater weight for male applicants. On the other hand, for female applicants, the presence of a full professor in the same university with the same family name as the candidate assumes greater weight than for male candidates.

## 1. Introduction

It is commonly understood that women are underrepresented in the research systems of many countries. In fact the data on national research staffing reveal a significant gap in the presence of women. Only four of 28 OECD nations[2] (Portugal, Estonia, Slovak Republic, Iceland) have a percentage of women greater than 40% in their national systems, and in none of these cases does female representation exceed 46% (OECD 2014). In the UK, women represent only 38.3% of total researchers, and in Italy only 34.5%. In France the share drops below 26.0%, and in Germany below 25%. Women researchers in Japan make up just 13.8% of the national staff. Although the four Nordic countries (Denmark, Finland, Norway, Sweden) are popularly considered progressive in women's rights, in these nations male scientists still outnumber their female colleagues two to one. The question that naturally arises is which of a series of factors could be the cause of this underrepresentation: lower numbers of women graduates; less interest among women for research activity; lesser scientific merit, and/or phenomena of gender bias in recruitment processes. The intention of the present work is to verify the hypothesis of the latter cause, thus contributing to a line of current studies on gender bias in the recruitment of academic staff (Zinovyeva and Bagues, 2015; Moss-Racusin et al., 2012: van den Brink et al., 2010; Wright et al., 2003; Fuchs et al., 2001). One branch of this literature demonstrates that discriminatory phenomena tend to appear when evaluations are not based on transparent criteria (Rees, 2004; Ziegler, 2001; Husu, 2000; Ledwith and Manfredi, 2000; Allen, 1988). In effect, academic recruitment is often described as an informal process, in which a few powerful professors promote or select new professors through mechanisms of cooptation (van den Brink et al., 2010; Husu, 2000; Fogelberg et al., 1999; Evans, 1995). A series of studies also show that women professors progress more slowly through academic ranks, tend not to attain important leadership roles, and earn less than men in comparable positions (Rotbart et al., 2012; Bilimoria and Liang, 2011; McGuire et al., 2004; Wright et al., 2003). The fact that women are underrepresented in decision-making positions appears to reduce probabilities for the recruitment and advancement of female candidates (Moss-Racusin et al., 2012; Corrice, 2009). De Paola and Scoppa (2015) find that "female candidates are less likely to be promoted when the committee is composed exclusively by males, while the gender gap disappears when the candidates are evaluated by a mixed sex committee". Zinovyeva and Bagues (2015, 2011) find that in competitions for full professor positions in Spain, evaluators tend to favor candidates who belong to their own academic network and are also of the same gender. In Italy, Abramo et al. (2015a) observed a moderate positive association between competitions with expected outcomes and the fact that the committee president was a woman. Bagues et al. (2014) estimated the causal effect of the gender composition of committees in the 2012 Italian competitions for qualification for associate and full professor positions. Differently from other studies, they found that each additional female evaluator decreases the success rate of female candidates by 2 percentage points.

The results of the preceding two studies on the Italian case are in our opinion not necessarily to be considered in disagreement. In fact the positive link between the increasing presence of women in the competition committees and the consideration of merit over favoritism (Abramo et al., 2015a) could coexist with the link between the increase in female representation in the committees and the diminution of success rates for women candidates (Bagues et al., 2014). The joint observations would indeed be logical, since it has been shown that female researchers are less productive than males in most disciplines (Larivière et al., 2013; Mauleón and Bordons, 2006; Xie and Shauman, 2004; Long, 1992; Fox, 1983), although gender differences are lessening over time (Frietsch et al., 2009; Abramo et al., 2009a; Alonso-Arroyo et al., 2007; Leahey, 2006; Xie and Shauman, 1998; Cole and Zuckerman, 1984). Moreover, the productivity gap is especially remarkable among top scientists (Abramo et al., 2009b; Bordons et al., 2003), who are those more likely to apply for higher academic positions.

[2] Data for the remaining 6 OECD nations (Australia, Canada, Israel, Mexico, New Zealand, and United States) are not available.

The Italian context is particularly suited to studies of gender bias given the high rate of favoritism in competitions for the public sector, which includes the university sphere. According to The Global Competitiveness Report 2013-2014 (Schwab, 2013), Italy ranks 126[th] out of 148 countries in favoritism in decisions of government officials. It is no surprise then that the nationally governed competitions for faculty positions have come under frequent fire, and that the Italian word "concorso" has gained international note for its implications of rigged competition, favoritism, nepotism and other unfair selection practices (Gerosa, 2001). Cases of favoritism in the faculty recruitment have been the subject of frequent media attention, and have even arrived before the courts (Zagaria, 2007; Perotti, 2008). A series of empirical studies have demonstrated that in Italy, scientific merit is not always the prevailing criteria for selection (Abramo et al., 2015b, 2014a, 2014b; Allesina, 2011; Durante et al., 2011, 2009).

In a context where favoritism is very diffuse it becomes easier to verify if, among those who are the subject of possible bias, there are differences in gender. In carrying out this verification, we distinguish between positive bias and negative bias in the competition outcomes. We then ask if the weights of the diverse factors that may determine the competition outcome indeed differ by gender. The two potential determinants of interest are: i) the scientific merit of the candidates, and ii) the possibilities for favoritism towards the candidates arising from factors of social proximity and research collaboration between the candidates and their evaluators, particularly involving the committee president.

The literature on gender discrimination and inequality in universities features several streams of activity, particularly qualitative research based on interviews (Bagilhole, 1993) and quantitative research based on surveys and questionnaires (McGuire et al., 2004; Wright et al., 2003), as well as analyses of academic faculties and their selection procedures (Rotbart et al., 2012; Moss-Racusin et al., 2012; Ceci and Williams, 2011). These studies identify the principle phenomena of gender discrimination as being: lesser probability for women to achieve promotion or tenure; lesser probability of obtaining leadership roles such as division head, department head, or dean; assignment of salaries that are lower than those of their male colleagues. However these studies, often focused on selected disciplines, lead to results that are difficult to generalize. Van den Brink et al. (2006) overcome these limits in their analysis of the reports from selection committees for 682 professorships in seven different disciplines for six large Dutch universities, over the period 1999-2003. However these authors themselves note a fundamental limitation in their work: the impossibility to elaborate a strict measurement of gender bias, given that they are unable to measure the quality of the applicants. The current study overcomes this limit by analyzing nearly the entirety of all scientific disciplines active in Italy, and evaluating the merit of the applicants through a bibliometric indicator of their research productivity.

The next section of the paper describes the recruitment process in Italian universities, particularly the measures adopted in 2008 for the recruitment of associate professors. Section 3 presents the dataset used for the analyses. Section 4 presents the results of our analyses on the gender bias, followed by the results of the regression analyses in Section 5. The work concludes with the authors' discussion.

## 2. Recruitment in Italian universities

The Italian Ministry of Education, Universities and Research (MIUR) recognizes a total of 96 universities as authorized to issue degrees. Sixty-seven of these are public universities, employing around 95% of all Italian faculty members.

In keeping with the so-called Humboldtian model of university policy, there are no "teaching-only" universities in Italy. All professors are required to carry out both research and teaching. Legislation includes a provision that each faculty member must provide a minimum of 350 hours of teaching per year. All new personnel enter the university system through public competitions, and career advancement can only proceed by further public competitions. Salaries are regulated at the centralized level and are calculated according to role (administrative, technical, or professorial), rank within role (for example assistant, associate or full professor) and seniority. None of a professor's salary depends on merit. Moreover, as in all Italian public administration, the dismissal of unproductive employees is unheard of.

The recruitment and advancement of professors is regulated by laws, which are overseen by the MIUR. There have been major reforms over recent years. Law 240 of 2010 introduced a double evaluation procedure for the selection of associate and full professors. The first level is a stage of national prequalification for the candidates, managed directly by the MIUR. A second stage of evaluations is managed by the individual universities, to then choose the prequalified individuals best suited to the specific needs of each institution. Prior to Law 240, the processes of recruitment and career advancement were in the hands of the individual universities, which were to follow procedures set by the national ministry. The last major set of competitions under the old system was held in 2008: the relevant data on these competitions, which are the context for the present work, is described in Section 3.

In the Italian university system all professors are classified in one and only field (Scientific Disciplinary Sector or SDS, 370 in all), grouped into disciplines (University Disciplinary Areas or UDAs, 14 in all)[3]. In both the new and old system, competitions for recruitment and advancement are organized at the SDS level. The 2008 competition procedures required appointment of committees to judge the curricula of the candidates. Each committee was to be composed of five full professors belonging to the SDS for which the position was open. One member, the president, was designated by the university holding the competition and the other four were drawn at random from a short list of other full professors in the national SDS. The short list was in turn established by a vote of all full professors in the national SDS.

The task of each committee was to provide a judgment of all candidates based on examination of their documented qualifications, and name at most two winners. The

---

university announcing the competition could then hire one of the two top finishers. The other top finisher remained eligible for hiring by any other university in the national system without further competition, at any time over the next five years.

In order to rationalize the process of the individual competitions over the entire system, the MIUR monitored and gathered the hiring proposals of the various universities and supported the evaluation procedures through information management systems intended to guarantee greater transparency. One of the ministry measures was to provide a Web portal[4] with all the basic information on the competition procedures, the posts available, the number of candidates for each competition, the scheduling of the procedures and final results (lists of winners, etc.). The transparency provisions, the nomination of a national committee of experts in the field, and the timely issue of regulations for the evaluation procedures were all intended to ensure efficiency in the selection process. In reality, the characteristics of Italian system – such as the generally strong inclination to favoritism, the structured lack of consequences for poor performance by research units, and the lack of incentive schemes for merit – undermined the credibility of selection procedures for the hiring and advancement of university professors. In a preceding work (Abramo et al., 2014b), we revealed several critical issues, particularly concerning unsuccessful candidates who remarkably outperformed the competition winners in terms of productivity over the subsequent triennium, as well as a number of competition winners who resulted as totally unproductive. An analysis of the individual competitions showed that almost half of them selected candidates who would go on to achieve below-median productivity in their field of reference over the subsequent period. In a subsequent study (Abramo et al., 2015b), we found that the most important determinant of a candidate's success was not his or her scientific merit, rather the number of their years of service in the same university as the committee president. In the current of these studies we now wish to investigate whether gender differences in discrimination and favoritism occurred in such competitions.

## 3. Dataset

In 2008, 1,232 competitions for associate professor positions were announced by a total of 74 Italian universities. The competitions concerned a total of 299 SDSs. At the end of all the selection processes, which lasted an average of over two years, the committees had named 2,339 winners from a total of 16,500 candidates[5]. The ratio of the number of competition winners to the size of the existing national associate professor faculty was 12.8. The competitions generally announced two winners (only 39 announced one winner).

To ensure the representativeness of publications as proxy of research output for the bibliometric assessment of the research merit of candidates, our analysis focuses only on the competitions in what we define the "bibliometric sectors", i.e. those SDSs where at least 50% of professors produced at least one publication indexed in the Web of Science™ (WoS) over the period 2004-2008. The bibliometric SDSs cover all the hard sciences and a few fields of Economics. For the observed period, there were 654 competitions that met such criteria, in a range of 193 SDSs. The only way to identify all the applicants in these competitions would be to read the minutes of each competition, as were generally published on-line by the individual universities. Given the prohibitive scope of such a task we have selected a further subset of 287 competitions was extracted from the population (44% of the total 654 in the hard sciences, in 124 SDSs). For this subset, the winners (550 in all) represent 22% of the total candidates (2,590). The rate of selection was more favorable for candidates who were incumbent assistant professors (532/2,314=23.0%) than it was for other individuals (18/276=6.5%). Due to the difficulties of authorship disambiguation, our research method is only able to measure the productivity of applicants who are already incumbent faculty members, thus our analysis of career advancement concentrates solely on the assistant professor candidates. In addition, for reasons of robustness, the measure of research productivity must be calculated over a sufficiently long period (Abramo et al., 2012a). Because of this, the analysis excludes assistant professors who entered faculty less than three years prior to the date of the competition. The dataset for the analysis is thus composed of 1,979 assistant professors, 473 of which were competition winners. Table 1 provides the characteristics of our dataset by UDA and its coverage with respect to overall competitions in the 193 SDSs of the hard sciences. On average there were nine participants per competition, of which eight were Italian-national academics. However the number of candidates shows significant variation (standard deviation 5.6), and 16 competitions involve 20 or more candidates. In the majority of competitions (263 out of 287) the committee designated two winners, with only 24 competitions resulting in a single winner.

*Table 1: Population subset selected for analysis (in parentheses the percentage with respect to the overall reference population by UDA)*

| UDA | Competitions | SDSs concerned | Winners | Academic winners with seniority ≥ 3 years |
|---|---|---|---|---|
| Mathematics and computer science | 26 (46%) | 7 (78%) | 50 (46%) | 45 (47%) |
| Physics | 19 (42%) | 5 (63%) | 37 (43%) | 30 (41%) |
| Chemistry | 25 (46%) | 8 (67%) | 47 (46%) | 44 (48%) |
| Earth sciences | 6 (30%) | 4 (33%) | 10 (27%) | 5 (17%) |
| Biology | 25 (34%) | 14 (74%) | 49 (34%) | 39 (31%) |
| Medicine | 62 (41%) | 32 (68%) | 116 (40%) | 87 (40%) |
| Agricultural and veterinary sciences | 15 (31%) | 11 (39%) | 27 (29%) | 26 (30%) |
| Civil engineering | 11 (42%) | 6 (86%) | 22 (43%) | 22 (46%) |
| Industrial and information engineering | 86 (60%) | 31 (74%) | 170 (62%) | 155 (62%) |
| Pedagogy and psychology | 5 (24%) | 3 (60%) | 8 (20%) | 7 (21%) |
| Economics and statistics | 7 (39%) | 3 (75%) | 14 (39%) | 13 (41%) |
| Total | 287 (44%) | 124 (64%) | 550 (43%) | 473 (44%) |

Table 2 presents the descriptive statistics concerning the candidates, by gender.

*Table 2: Descriptive statistics for the candidates involved in the dataset of competitions, by gender*

| | | | | | Candidates per competition | | | |
| --- | --- | --- | --- | --- | --- | --- | --- | --- |
| | | Winners | Non winners | Total | Average | Median | Std Dev | Max |
| | Female | 162 | 586 | 748 | 3 | 2 | 2.6 | 14 |
| Total candidates | Male | 388 | 1,454 | 1,842 | 6 | 5 | 4.7 | 24 |
| | Total | 550 | 2,040 | 2,590 | 9 | 8 | 5.6 | 29 |
| | Female | 154 | 518 | 672 | 2 | 2 | 2.5 | 14 |
| Academics | Male | 378 | 1,264 | 1,642 | 6 | 5 | 4.3 | 23 |
| | Total | 532 | 1,782 | 2,314 | 8 | 7 | 5.4 | 28 |
| | Female | 8 | 68 | 76 | 0 | 0 | 0.5 | 2 |
| Others | Male | 10 | 190 | 200 | 1 | 0 | 1.0 | 5 |
| | Total | 18 | 258 | 276 | 1 | 1 | 1.2 | 6 |
| Academics with seniority ≥ 3 years | Female | 141 | 467 | 608 | 2 | 1 | 2.3 | 12 |
| | Male | 332 | 1,039 | 1,371 | 5 | 4 | 3.8 | 22 |
| | Total | 473 | 1,506 | 1,979 | 7 | 6 | 4.6 | 26 |
| Academics with seniority < 3 years | Female | 13 | 51 | 64 | 0 | 0 | 0.6 | 3 |
| | Male | 46 | 225 | 271 | 1 | 1 | 1.2 | 5 |
| | Total | 59 | 276 | 335 | 1 | 1 | 1.3 | 6 |

## 4. Bias in academic recruitment

The selection committees judge the applicants for academic positions based on both quantitative and qualitative criteria. The committees are free to define the evaluation models they will apply, relative to the scientific sector and academic rank of concern. Our own investigation of the possible cases of unfair evaluation is based only on the research performance of the candidates. Furthermore, The bibliometric assessment of research performance by quantity and quality of output neglects other attributes of the scientists' activities, for example the ability to manage research teams, to attract funds, their activities in consulting, teaching, editorial work, outreach, and so on. Still, common sense would lead one to believe that there is a strong correlation between research productivity and all other dimensions of scientific merit. For example, Marsh and Hattie (2002), Elton (2001), and Hattie and Marsh (1996) show evidence of a positive correlation between research productivity and teaching effectiveness. The assessment of potential cases of discrimination and favoritism in academic recruitment is therefore subject to a certain degree of uncertainty. Furthermore, like all measures, the FSS measurement itself embeds some degree of uncertainty. In the following, when we refer to bias in recruitment, we always imply the embedded uncertainty. For simplicity of language we will generally refer to negative bias as "discrimination", and to positive bias as "favoritism". Because of the limits and assumptions embedded in the methodology and performance indicator applied to identify possible cases of bias in recruitment, the usual warnings in the interpretations of results apply.

### 4.1 Negative bias

We define negative bias (discrimination) as a situation where a non-winner candidate is observed to have scientific merit notably higher than that of at least one the competition winners, and not less than that of the other non-winner candidates. As the indicator of the scientific merit of a candidate we use their research productivity, quantified by an indicator named Fractional Scientific Strength, or FSS, which embeds both the number of publications and their field-normalized citations (for a detailed description of the index and the underlying theory, see Appendix).

Since the committees also apply criteria other than research productivity in making their selections, we assume that within a difference of 20 FSS percentiles[6] the other dimensions of merit could compensate for the difference in research productivity. This threshold appears reasonable, since it is coherent to expect a positive relation between research performance and the other variables of academic merit. Thus if a candidate places 20 percentiles above another, according to the convention we adopt, he or she surely has more merit. If the difference is less than 20, we cannot affirm that there is a difference in merit. Since it is possible that all applicants in a competition would be of little merit, in order for a candidate to be defined as "discriminated against" it is also necessary that their FSS must in all cases be higher than the median of the national performance distribution of all their colleagues of the same rank and SDS. In this regard, we also recall the committees would have been free not to name any winners.

In summary, the conditions necessary for a candidate to be defined as subject to negative bias are:

i.   There must be a positive difference of 20 percentiles (national ranking of assistant professors, by FSS) between the non-winner candidate and the worst of the winners;

ii.  The FSS of the candidate must not be less than the median of the national distribution of assistant professors in the SDS;

iii. There must be a negative difference of performance of not more than 20 percentiles between the non-winning candidate and the best of those that satisfy the first two conditions.

Given the above conditions there could more than one subject of discrimination per competition.

To ensure a robust analysis of bibliometric productivity (Abramo et al., 2012a), of the 287 competition analyzed in detail up to this point we now further exclude those competitions (44 out of 287) lacking at least a winner and a non-winning participant with at least three years in a faculty position over the 2004-2008 period. Given the exclusions, we reduce the number of competitions observed to 243.

We now present the results of the gender differences in discrimination, analyzed from two points of view: the numerosity of the subjects of discrimination and the extent of the discrimination.

We first measure the FSS of all candidates and assistant professors in the SDSs where the 243 competitions were launched. On the basis of the defined conditions we then identify the candidates that were subject to discrimination during the judgment of the competition. Finally we investigate the gender differences of such discrimination, by UDA (Table 3).

From the analysis it emerges that there are a total of 323 cases of discrimination out of 1,883 applicants, relative to 422 winners. Of the subjects of discrimination, 24.1% are women against 75.9% that are men, compared to 30.6% of applicants being female and 64.9% being male. Student's t-test confirms that female applicants are significantly less subject to discrimination than males (p-value<0.01).

---

[6] As explained in detail in the Appendix, the FSS percentile refers to the distribution of productivity of all the national assistant professors in the same SDS.

**_Table 3: Candidates biased against, applicants, and correlation between FSS and competition outcome, by gender and UDA (% in brackets)_**

| UDA | Biased against | | | Applicants | | | Correlation FSS-competition outcome | | |
|---|---|---|---|---|---|---|---|---|---|
| | F | M | Tot | F | M | Tot | F | M | Tot |
| Mathematics and computer science | 7 (21.2) | 26 (78.8) | 33 | 101 (35.3) | 185 (64.7) | 286 | 0.330*** | 0.255*** | 0.284*** |
| Physics | 5 (16.1) | 26 (83.9) | 31 | 40 (23.7) | 129 (76.3) | 169 | 0.007 | 0.187* | 0.123 |
| Chemistry | 9 (31.0) | 20 (69.0) | 29 | 65 (46.1) | 76 (53.9) | 141 | 0.108 | 0.347** | 0.207* |
| Earth sciences | - | - | - | 5 (22.7) | 17 (77.3) | 22 | 0.409 | 0.298 | 0.313 |
| Biology | 15 (53.6) | 13 (46.4) | 28 | 110 (65.5) | 58 (34.5) | 168 | -0.019 | -0.090 | -0.038 |
| Medicine | 12 (19.0) | 51 (81.0) | 63 | 65 (21.9) | 232 (78.1) | 297 | -0.013 | 0.076 | 0.062 |
| Agricultural and veterinary sciences | 4 (44.4) | 5 (55.6) | 9 | 15 (38.5) | 24 (61.5) | 39 | -0.160 | -0.146 | -0.133 |
| Civil engineering | 2 (8.0) | 23 (92.0) | 25 | 24 (21.8) | 86 (78.2) | 110 | 0.443** | -0.004 | 0.097 |
| Industrial and information engineering | 16 (17.2) | 77 (82.8) | 93 | 113 (19.6) | 465 (80.4) | 578 | 0.222** | 0.134*** | 0.144*** |
| Pedagogy and psychology | 3 (60.0) | 2 (40.0) | 5 | 11 (55.0) | 9 (45.0) | 20 | 0.098 | 0.030 | 0.055 |
| Economics and statistics | 5 (71.4) | 2 (28.6) | 7 | 27 (50.9) | 26 (49.1) | 53 | 0.238 | 0.182 | 0.216 |
| Total | 78 (24.1) | 245 (75.9) | 323 | 576 (30.6) | 1,307 (69.4) | 1,883 | 0.140*** | 0.101*** | 0.113*** |

*Statistical significance: *p-value <0.10, **p-value <0.05, ***p-value <0.01.*
*Statistical significance level adjusted using Bonferroni corrections*

Deepening the examination to the level of the UDAs, the analyses show that the incidence of discriminated women out of the total of subjects of discrimination is the highest in Biology (53.6%), where we also observe the highest percentage of female applicants (65.5%). Civil engineering has the lowest percentage of female subjects of discrimination out of the total (8%). The gender differences in terms of concentration of those discriminated in relation to concentration of applicants results as significant in three UDAs out of 10[7]: men result as more discriminated in Mathematics and computer science (p-value = 0.053), Civil engineering (p-value = 0.058), and Chemistry (p-value = 0.069); women do not result as more subject to discrimination in any of the UDAs.

Table 3 also presents the correlation between FSS and competition outcome (competition outcome = 1, if the applicant wins the competition; 0, otherwise), by gender and UDA. In the case of women, Civil engineering has the highest Pearson correlation (0.443). In the case of the men, the highest correlation is observed in Chemistry (0.347).

For each of the 323 applicants subject to discrimination we measure the level of discrimination in percentiles of FSS, calculated as follows:

$$D = FSS_d - (FSS_w + 20)$$

[1]

Where:
D = level of discrimination;
$FSS_d$ = FSS percentile of the candidate discriminated;
$FSS_w$ = FSS percentile of the worst of the winners.

Table 4 shows the average, maximum and standard deviation of the level of discrimination, by gender and UDA.

The general analysis shows that the highest discrimination occurs in Civil engineering (average discrimination = 35.4 FSS percentiles), the lowest in Economics and statistics (average discrimination = 7.4 FSS percentiles). The average discrimination for women candidates is 21.0 FSS percentiles; that for male candidates is 21.1 percentiles. Student's t-test shows that the averages of female and male levels of discrimination are not significantly different (p-value=0.459).

In effect, the gender differences in the level of discrimination result as significant in only one UDA out of 10, which is Pedagogy and psychology (34.5 for women vs. 1.9 for men, p-value = 0.023).

---

[7] For Earth sciences, since there are no subjects of discrimination it is not possible to deepen the analyses for the gender differences in this regard.

Table 4: Levels of negative bias by gender and UDA, measured in percentiles of FSS

| UDA | F | | | M | | | Tot | | |
|---|---|---|---|---|---|---|---|---|---|
| | Avg | Std dev. | Max | Avg | Std dev. | Max | Avg | Std dev. | Max |
| Mathematics and computer science | 11.9 | 9.0 | 27.0 | 12.7 | 10.9 | 35.6 | 12.5 | 10.4 | 35.6 |
| Physics | 16.3 | 10.5 | 32.8 | 20.5 | 15.9 | 52.1 | 19.8 | 15.1 | 52.1 |
| Chemistry | 20.5 | 16.2 | 47.5 | 18.0 | 13.2 | 42.2 | 18.8 | 14.0 | 47.5 |
| Earth sciences | - | - | - | - | - | - | - | - | - |
| Biology | 23.1 | 14.5 | 59.1 | 19.0 | 15.0 | 57.2 | 21.2 | 14.6 | 59.1 |
| Medicine | 21.6 | 12.1 | 37.9 | 22.9 | 12.4 | 50.6 | 22.6 | 12.3 | 50.6 |
| Agricultural and veterinary sciences | 25.8 | 32.5 | 73.2 | 37.9 | 19.0 | 62.7 | 32.5 | 24.8 | 73.2 |
| Civil engineering | 36.2 | 47.9 | 70.1 | 35.4 | 31.2 | 79.2 | 35.4 | 31.5 | 79.2 |
| Industrial and information engineering | 23.1 | 18.8 | 58.3 | 19.7 | 16.6 | 73.7 | 20.3 | 16.9 | 73.7 |
| Pedagogy and psychology | 34.5 | 30.3 | 58.5 | 1.9 | 0.2 | 2.0 | 21.5 | 27.9 | 58.5 |
| Economics and statistics | 6.6 | 8.1 | 19.8 | 9.6 | 0.6 | 10.1 | 7.4 | 6.8 | 19.8 |
| Total | 21.0 | 17.5 | 73.2 | 21.1 | 17.7 | 79.2 | 21.1 | 17.6 | 79.2 |

## 4.2 Positive bias

If there is negative bias (discrimination) against some of the candidates in a competition then at least one winner must have undergone positive bias and been favored. In this section we analyze the gender differences among those favored in the competitions. We define favoritism as the situation where the winner of a competition has a lower scientific merit than at least one non-winning candidate, or in any case lower than what we would have expected of a winner with respect to all the other colleagues of the same academic rank and SDS.

As in the case of the calculation of discrimination, since FSS is not the only criterion for selection in the competitions, we assume that within 20 percentile points the other variables for merit could compensate for the difference in research productivity. Thus, if a candidate places 20 percentiles below another for FSS, he or she is surely less worthy. If the difference is less than 20, this cannot be concluded.

Summarizing, the sufficient conditions for identifying a candidate as favored in a competition are either of the following:

i. There is a negative difference in performance of not less than 20 percentile points (national ranking of assistant professors, by FSS) between the winning candidate and the best of the non-winning applicants;

ii. The FSS of the winner is less than the national median of the assistant professors in the SDS.

As much as the occurrence of the second condition seems unlikely, we observe that: in 78 competitions at least one winner showed performance less than the national median; in 13 competitions a candidate resulted as selected while satisfying the second condition only; in one competition all the candidates had productivity below the national median, but two winners were still chosen. The question that naturally arises is not why assistant professors with a poor scientific portfolio would have entered the competitions, but why those with a high profile did not. Among those familiar with the culture and the practices of favoritism in the Italian university environment it is well known that, for purposes of "taking turns" in sharing out of positions, full professors will often place pressure on worthy assistant professors in their universities to hold back from entering competitions, in order not to create problems at the moment when the committees move to select the predetermined individual for that occasion.

For the same 243 competitions examined in the analysis of discrimination, we thus identify those candidates that were favored in the judgment stage and analyze the potential gender differences that may have occurred in such favoritism (Table 5).

From the analysis there emerge a total of 186 favored candidates out of 422 winners and 1,833 applicants. Some 32.8% of the total favored candidates are women, against 67.2% men; compared to 30.6% of the applicants being of female gender and 64.9% male. Student's t-test does not indicate gender differences among favored candidates (p-value=0.246).

***Table 5: Candidates benefitting from positive bias and applicants, by gender and UDA (% in brackets)***

| UDA | Favored candidates | | | Applicants | | |
|---|---|---|---|---|---|---|
| | F | M | Tot | F | M | Tot |
| Mathematics and computer science | 3 (23.1) | 10 (76.9) | 13 | 101 (35.3) | 185 (64.7) | 286 |
| Physics | 6 (33.3) | 12 (66.7) | 18 | 40 (23.7) | 129 (76.3) | 169 |
| Chemistry | 10 (52.6) | 9 (47.4) | 19 | 65 (46.1) | 76 (53.9) | 141 |
| Earth sciences | - | - | - | 5 (22.7) | 17 (77.3) | 22 |
| Biology | 10 (50) | 10 (50) | 20 | 110 (65.5) | 58 (34.5) | 168 |
| Medicine | 10 (28.6) | 25 (71.4) | 35 | 65 (21.9) | 232 (78.1) | 297 |
| Agricultural and veterinary sciences | 3 (50) | 3 (50) | 6 | 15 (38.5) | 24 (61.5) | 39 |
| Civil engineering | 1 (16.7) | 5 (83.3) | 6 | 24 (21.8) | 86 (78.2) | 110 |
| Industrial and information engineering | 15 (24.6) | 46 (75.4) | 61 | 113 (19.6) | 465 (80.4) | 578 |
| Pedagogy and psychology | 3 (60) | 2 (40) | 5 | 11 (55) | 9 (45) | 20 |
| Economics and statistics | 0 (0) | 3 (100.0) | 3 | 27 (50.9) | 26 (49.1) | 53 |
| Total | 61 (32.8) | 125 (67.2) | 186 | 576 (30.6) | 1,307 (69.4) | 1,883 |

Deepening the analysis by UDA, we observe that the incidence of favored women in the total of favored candidates is highest in Pedagogy and psychology (60.0%). Civil engineering has the lowest percentage of favored women in the total (16.7%). Gender differences among favored winners result as significant in two UDAs out of 10[8]. In both cases it is the men that result as more favored than women: in Economics and Statistics (p-value=0.036) and in Biology (p-value=0.061).

Table 6 shows the level of favoritism in percentiles of FSS, calculated as follows:

$$F = FSS_a - (FSS_f + 20)$$

[2]

Where:

F = level of favoritism;
$FSS_a$ = percentile of FSS of the best non-winner candidate;
$FSS_f$ = percentile of FSS of the favored candidate.

The analyses show the average, the standard deviation and the maximum level of favoritism, by gender and UDA. The general analysis shows that the highest level of favoritism is in Civil engineering (average favoritism = 35.8 percentiles of FSS), the lowest is in Economics and statistics (average = 10.5 percentiles of FSS). The average favoritism for the women candidates equals 19.8 percentiles of FSS; that for the men candidates equals 19.5 percentiles of FSS. Student's t-test shows that the averages of female and male favoritism or not significantly different (p-value=0.456).

In the analyses at the more detailed level, the gender differences in level of favoritism result as not significant for all the UDAs.

---

[8] For Earth sciences, since there are no subjects of favoritism it is not possible to deepen the analyses for the gender differences in this regard.

*Table 6: Level of positive bias by gender and UDA, measured in percentiles of FSS*

| UDA | F | | | M | | | Tot | | |
|---|---|---|---|---|---|---|---|---|---|
| | Avg | Std dev. | Max | Avg | Std dev. | Max | Avg | Std dev. | Max |
| Mathematics and computer science | 17.4 | 15.1 | 27.0 | 9.7 | 10.7 | 35.6 | 11.5 | 11.6 | 35.6 |
| Physics | 20.5 | 19.2 | 52.1 | 17.0 | 15.0 | 40.2 | 18.2 | 16.1 | 52.1 |
| Chemistry | 18.3 | 15.5 | 47.5 | 17.1 | 13.3 | 39.6 | 17.7 | 14.0 | 47.5 |
| Earth sciences | - | - | - | - | - | - | - | - | - |
| Biology | 17.5 | 10.0 | 35.0 | 21.3 | 18.3 | 59.1 | 19.4 | 14.4 | 59.1 |
| Medicine | 20.6 | 13.2 | 47.9 | 20.9 | 12.7 | 50.6 | 20.8 | 12.6 | 50.6 |
| Agricultural and veterinary sciences | 14.9 | 8.2 | 23.0 | 46.4 | 37.8 | 73.2 | 30.6 | 29.9 | 73.2 |
| Civil engineering | 21.4 | - | 21.4 | 38.7 | 33.7 | 79.2 | 35.8 | 31.0 | 79.2 |
| Industrial and information engineering | 21.8 | 16.7 | 60.8 | 18.7 | 18.5 | 73.7 | 19.6 | 18.0 | 73.7 |
| Pedagogy and psychology | 24.6 | 30.0 | 58.5 | 2.0 | - | 2.0 | 18.9 | 26.9 | 58.5 |
| Economics and statistics | - | - | - | 10.5 | 9.2 | 19.8 | 10.5 | 9.2 | 19.8 |
| Total | 19.8 | 14.8 | 60.8 | 19.5 | 18.2 | 79.2 | 19.6 | 17.1 | 79.2 |

## 5. Statistical Analysis

For the competitions of our dataset, we formulate a statistical model that links the competition outcome to the possible determinants, as described below.

The dependent variable, the competition outcome, is a Boolean type variable with value of 1 in the case that the applicant wins, or 0 otherwise. The eight independent variables are: the parental link between applicant and full professors in the same university (NE); the career years that an applicant has spent in the same university and same SDS as the committee president (CP); the career years that an applicant has spent in the same university and the same SDS as other committee members (CE); the percentage of the president's publications coauthored with the candidate (PP); the number of other committee members with which the applicant has co-authored publications (PE); the applicant's scientific productivity (FSS) for the five years 2004-2008, as proxy of scientific merit; the agreement between the gender of the applicant and the gender of the committee president (SP); and the agreement between the gender of the applicant and at least three committee members (president included) (SE).

To study the effect of gender on the eight independent variables which determine the competition outcome, we introduce the Boolean independent variable G, whose value is 1 in the case that the applicant is female, and 0 otherwise. In addition to the eight independent variables presented above we introduce in the model the same variables each multiplied by G.

As the basis of the statistical model we choose the logistic regression function (linearized by the logit function), which is particularly suited for modeling dichotomous dependent variables. Formally, the statistical model is described as:

$$logit(p) = \beta_0 + \beta_1 G + \beta_2 FSS + \beta_3 G \cdot FSS + \beta_4 NE + \beta_5 G \cdot NE + \beta_6 CP + \beta_7 G \cdot CP$$
$$+ \beta_8 CE + \beta_9 G \cdot CE + \beta_{10} PP + \beta_{11} G \cdot PP + \beta_{12} PE + \beta_{13} G \cdot PE$$
$$+ \beta_{14} SP + \beta_{15} G \cdot SP + \beta_{16} SE + \beta_{17} G \cdot SE)$$

[3]

Where:

$$logit(p) = \log \frac{p(E)}{1 - p(E)}$$
[4]

$E$ = competition outcome: 1, if the applicant wins the competition; 0, otherwise;

$p(E)$ = probability of event E;

$\beta$ = generic regression coefficient;

G = 1, if the applicant is female; 0, if the applicant is male.

FSS = applicant's research productivity over the period 2004-2008, expressed on a 0-100 percentile scale;

NE = 1, if the applicant and a full professor in the same university have the same family name; 0, otherwise.

CP = applicant's career years in the same university and same SDS as the committee president over the period 2001-2010.

CE = applicant's career years in the same university and the same SDS as the other evaluation committee members over the period 2001-2010.

PP = percentage of committee president's publications in co-authorship with the candidate over the period 2001-2010.

PE = number of other committee members with which the applicant has co-authored publications over the period 2001-2010.

SP = 1, if the applicant has the same gender as the committee president; 0, otherwise.

SE = 1, if the applicant has the same gender as at least three committee members (president included); 0, otherwise.

## 5.1 Descriptive statistics

Prior to applying the statistical model we present the descriptive statistics for the basic variables in Table 7, distinguishing by applicant gender. For each variable we show the average, standard deviation (SD) and the maximum value occurring for the winners, non-winners and total applicants in the dataset.

In the case of the female applicants, the winners' scientific performance is on average higher than that of non-winners (65.13 for winners versus 57.14 for non-winners). The average number of years that the female applicant spent in the same university as the committee president (CP) is 1.99; for winners this figure rises to 4.16 and for non-winners it drops to 1.34. For the set of all female applicants, the average number of years spent in the same university as the other committee members (CE) is 1.50, compared to 1.13 for the winners and 1.62 for non-winners. Concerning co-authored publications, on average the full set of female participants contribute to 2.45% of the president's scientific production; winners contribute to 8.10% and non-winners to 0.75%. Some 14% of the female applicants are of the same gender as the committee president; this percentage rises to 16% in the case of the winners and remains at 14% for the non-winners.

In the case of male applicants, once again the winners' scientific performance is on average higher than that of non-winners (69.64 for winners versus 63.80 for non-winners), although the difference is less than for females (65.13 for winners versus 57.14 for non-winners). The average number of years that the male applicant spent in the same university as the committee president is 2.16; for winners this figure rises to 4.31 and for non-winners it drops to 1.47. For the set of all male applicants, the average number of years spent in the same university as the other evaluators is 1.17, compared to 1.27 for the winners and 1.14 for the non-winners. Concerning research collaboration, on average the full set of male participants contribute to 2.34% of the president's scientific production; winners contribute to 6.55% and non-winners to 0.99%. Some 91% of the male applicants are the same gender as the committee president; this percentage equals 95% in the case of winners and 90% in the case non-winners. Finally, we note that concerning the variable NE (signal of possible cases of nepotism), an average of 4% of female applicants have the same family name as a full professor in the university holding the competition. This percentage increased to 7% in the case of female winners and drops to 3% in the case of female non-winners. In the case of male applicants this percentage equals 4% both for the winners and non-winners.

*Table 7: Descriptive statistics for logistic regression variables, by applicant gender*

| | Female Applicants | | | | | | | | |
|---|---|---|---|---|---|---|---|---|---|
| Var. | Winners | | | Non winners | | | Total | | |
| | Avg | SD | Max | Avg | SD | Max | Avg | SD | Max |
| FSS | 65.13 | 21.03 | 99.12 | 57.14 | 24.65 | 100 | 58.99 | 24.08 | 100 |
| NE | 0.07 | 0.26 | 0 | 0.03 | 0.17 | 1 | 0.04 | 0.19 | 1 |
| CP | 4.16 | 4.37 | 10 | 1.34 | 3.15 | 10 | 1.99 | 3.67 | 10 |
| CE | 1.13 | 3.59 | 20 | 1.62 | 4.01 | 21 | 1.50 | 3.92 | 21 |
| PP | 8.10 | 18.55 | 71.88 | 0.75 | 4.85 | 68.89 | 2.45 | 10.35 | 71.88 |
| PE | 0.13 | 0.42 | 3 | 0.09 | 0.30 | 2 | 0.10 | 0.33 | 3 |
| SP | 0.16 | 0.36 | 1 | 0.14 | 0.34 | 1 | 0.14 | 0.35 | 1 |
| SE | 0.08 | 0.27 | 1 | 0.08 | 0.27 | 1 | 0.08 | 0.27 | 1 |

| | Male Applicants | | | | | | | | |
|---|---|---|---|---|---|---|---|---|---|
| Var. | Winners | | | Non winners | | | Total | | |
| | Avg | SD | Max | Avg | SD | Max | Avg | SD | Max |
| FSS | 69.64 | 24.20 | 100 | 63.80 | 24.93 | 99.73 | 65.21 | 24.87 | 100 |
| NE | 0.04 | 0.20 | 1 | 0.04 | 0.20 | 1 | 0.04 | 0.20 | 1 |
| CP | 4.31 | 4.37 | 10 | 1.47 | 3.15 | 10 | 2.16 | 3.69 | 10 |
| CE | 1.27 | 3.44 | 20 | 1.14 | 3.03 | 20 | 1.17 | 3.13 | 20 |
| PP | 6.55 | 16.32 | 91.18 | 0.99 | 6.76 | 97.36 | 2.34 | 10.23 | 97.36 |
| PE | 0.12 | 0.35 | 2 | 0.08 | 0.28 | 2 | 0.09 | 0.30 | 2 |
| SP | 0.95 | 0.21 | 1 | 0.90 | 0.30 | 1 | 0.91 | 0.28 | 1 |
| SE | 0.97 | 0.16 | 1 | 0.97 | 0.16 | 1 | 0.97 | 0.16 | 1 |

## 5.2 Correlation Analysis

Table 8 presents the correlations between the basic regressors, comparing the case of the female applicants to that of the male applicants. In the case of the women, the Pearson correlation analysis indicates that the highest correlations are between CP and PP, at 0.386, and between CE and PE, at 0.361. This is in line with what we expect, since scientists in the same university and SDS would tend to cooperate in shared research work. The test of multicollinearity between the variables shows that the average VIF (Variance inflation factor) is 1.10. In the case of the male applicants, the Pearson correlation analysis indicates that the highest correlations are once again between PP and CP (0.356) and PE and CE (0.335). The test of multicollinearity shows that the average VIF is 1.09.

Thus from the Pearson correlation analysis and the test of multicollinearity between variables, for both cases of male and female applicants it emerges that the hypothesis of independence between the variables can be considered valid.

*Table 8: Correlation among variables, by applicant gender*

| | E | FSS | NE | CP | CE | PP | PE | SP | SE |
|---|---|---|---|---|---|---|---|---|---|
| | | | | Female applicants | | | | | |
| E | 1 | | | | | | | | |
| FSS | 0.140*** | 1 | | | | | | | |
| NE | 0.089 | 0.013 | 1 | | | | | | |
| CP | 0.326*** | 0.054 | -0.020 | 1 | | | | | |
| CE | -0.052 | -0.049 | -0.022 | -0.188*** | 1 | | | | |
| PP | 0.300*** | 0.026 | 0.055 | 0.386*** | -0.091 | 1 | | | |
| PE | 0.060 | -0.028 | -0.035 | -0.137** | 0.361*** | -0.071 | 1 | | |
| SP | 0.023 | 0.097 | -0.058 | 0.055 | 0.024 | 0.026 | -0.079 | 1 | |
| SE | -0.002 | -0.048 | 0.003 | -0.021 | 0.028 | 0.071 | -0.014 | 0.039 | 1 |
| | | | | Male applicants | | | | | |
| | E | FSS | NE | CP | CE | PP | PE | SP | SE |
| E | 1 | | | | | | | | |
| FSS | 0.101*** | 1 | | | | | | | |
| NE | 0.002 | -0.027 | 1 | | | | | | |
| CP | 0.330*** | -0.031 | -0.010 | 1 | | | | | |
| CE | 0.018 | -0.047 | 0.001 | -0.210*** | 1 | | | | |
| PP | 0.233*** | 0.051 | -0.017 | 0.356*** | -0.072 | 1 | | | |
| PE | 0.065 | 0.033 | -0.035 | -0.070 | 0.335*** | -0.019 | 1 | | |
| SP | 0.081* | -0.018 | 0.027 | 0.038 | -0.002 | 0.052 | 0.047 | 1 | |
| SE | -0.000 | -0.028 | 0.012 | 0.014 | 0.020 | 0.011 | -0.027 | -0.036 | 1 |

*Statistical significance: *p-value <0.10, **p-value <0.05, ***p-value <0.01*
*Statistical significance level adjusted using Bonferroni corrections*

### 5.3 The Logistic Regression Model

Table 9 presents the logistic regression results predicting the competition outcomes.

The odds ratio for the competition outcomes (i.e. probability of winning the competition relative to the probability of not winning) is formalized as:

$$\frac{p(E)}{1-p(E)} = \exp(-3.282 + 0.404\,G + 0.012\,FSS + 0.004\,G \cdot FSS + 0.127\,NE + 0.983\,G$$
$$\cdot NE + 0.188\,CP - 0.028\,G \cdot CP + 0.067\,CE - 0.085\,G \cdot CE + 0.024\,PP$$
$$+ 0.020\,G \cdot PP + 0.435\,PE + 0.533\,G \cdot PE + 0.691\,SP - 0.646\,G \cdot SP$$
$$- 0.032\,SE + 0.011\,G \cdot SE)$$

[5]

The value $e^b$, calculated for each potential explanatory variable, represents the odds ratio (OR)[9] in Table 9. The values calculated for standardized $b$ (last column, Table 9) permit comparison of the effects of the variables measured in different metrics.

The data indicate that the factor having the greatest influence on the competition outcomes ($b$Std$_{CP}$=0.693, p-value<0.01) seems to be the number of the applicant's years in the same university and same SDS as the committee president (CP). In particular, every unit increase in the number of career years shared with the president increases the odds of success by a factor of 1.207. Gender differences are not significant for this regressor.

The applicant's scientific productivity (FSS) also has remarkable bearing on the competition results ($b$Std$_{PP}$=0.304, p-value<0.01): every percent increase in FSS increases the odds of success by a factor of 1.012. The effect of the productivity on the outcome does not vary with gender.

---

[9] The "odds ratio" is used in statistics to quantify how strongly the presence or absence of property A is associated with the presence or absence of property B, in a given population. In our case, where OR equals 1 the associated explanatory variable would have no effect on the dependent variable, i.e. on competition outcome.

Like FSS, the co-authorship of publications with the committee president (PP) also has a remarkable weight ($b$Std$_{FSS}$=0.246, p-value<0.01), with every unit increase in the PP increasing the odds of success by a factor of 1.024. Here too, no gender differences occur. Our analysis does not show significant results concerning the weight of PE (i.e. the variable concerning shared research work with other committee members).

The applicant's career years in the same university as the other (non-president) selection committee members (CE) has a notable and significant bearing ($b$Std$_{CE}$ = 0.228; p-value<0.01). Here, gender differences are significant. CE has a greater weight on the outcome for male candidates ($b$Std$_{G*CE}$ = -0.194, p-value=0.046).

The SP variable (concerning agreement between the gender of the candidate and that of the committee president) also seems to have a certain weight (p-value<0.01)[10], which results greater for male candidates ($b_{G*SP}$ = -0.194, p-value=0.068).

The presence of a full professor in the same university with the same family name as the candidate (NE) has no significant effect on the outcome, although such effect is greater for women ($b_{G*NE}$ = 0.983, p-value=0.090).

Our analysis does not show significant results concerning the weight of SE and G (the variables concerning genders of committee members and applicant).

***Table 9: Logistic regression results predicting competition outcomes***

| | $b$ | OR | Std Err | Z | p>\|z\| | $b_{StdX}$ |
|---|---|---|---|---|---|---|
| G | 0.404 | 1.498 | 0.528 | 0.77 | 0.444 | 0.186§ |
| FSS | 0.012*** | 1.012 | 0.003 | 4.05 | 0.000 | 0.304 |
| G*FSS | 0.004 | 1.004 | 0.006 | 0.72 | 0.472 | 0.120 |
| NE | 0.127 | 1.135 | 0.330 | 0.39 | 0.699 | 0.025§ |
| G*NE | 0.983* | 2.672 | 0.580 | 1.69 | 0.090 | 0.108 |
| CP | 0.188*** | 1.207 | 0.018 | 10.28 | 0.000 | 0.693 |
| G*CP | -0.028 | 0.972 | 0.033 | -0.84 | 0.402 | -0.062 |
| CE | 0.067*** | 1.069 | 0.022 | 3.05 | 0.002 | 0.228 |
| G*CE | -0.085** | 0.919 | 0.043 | -1.99 | 0.046 | -0.194 |
| PP | 0.024*** | 1.024 | 0.009 | 2.65 | 0.008 | 0.246 |
| G*PP | 0.020 | 1.020 | 0.017 | 1.18 | 0.236 | 0.117 |
| PE | 0.435 | 1.545 | 0.281 | 1.55 | 0.121 | 0.135 |
| G*PE | 0.533 | 1.704 | 0.423 | 1.26 | 0.207 | 0.100 |
| SP | 0.691*** | 1.996 | 0.262 | 2.63 | 0.008 | 0.324§ |
| G*SP | -0.646* | 0.524 | 0.354 | -1.83 | 0.068 | -0.132§ |
| SE | -0.032 | 0.969 | 0.247 | -0.13 | 0.897 | -0.015§ |
| G*SE | 0.011 | 1.011 | 0.500 | 0.02 | 0.982 | 0.002§ |
| Constant | -3.282*** | - | 0.420 | -7.81 | 0.000 | |

§ *In this case the standardized coefficient is not considered because the explanatory variable is binary.*
*Dependent variable: competition outcome; method of estimation: logistic regression; b = raw coefficient; OR= Odds Ratio (exp b); z = z-score for test of b=0; p>|z| = p-value for z-test; bStd$_X$= X standardized coefficient.*
*Statistical significance: \*p-value <0.10, \*\*p-value <0.05, \*\*\*p-value <0.01.*
*Number of observations = 1,979.*
*Chi2 test for joint significance of interaction terms: chi2(17) = 239.85; Prob> chi2 = 0.0000.*
*Log likelihood = -940.1292; Pseudo R2 = 0.1362; Std Err. adjusted for 287 clusters (competitions).*
*Log likelihood = -940.1292; Pseudo R2 = 0.1362; Std Err. adjusted for 287 clusters (competitions).*

---

[10] Note that for NE, SP and SE the standardized coefficients lose meaning since the explanatory variables are binary.

## 6. Conclusions

In the current economic system, organizations compete at the worldwide level to attract and recruit the best talent possible. In this scenario, phenomena of a priori discrimination (e.g. for gender, age, culture, ethnicity or religion) imply a loss of effectiveness and efficiency in the human resources selection system, and negatively affect the organization's competitiveness. In higher education systems the avoidance of discriminatory phenomena becomes still more important, given the role that universities plays in supporting industrial competitiveness, socio-economic development and social mobility. In competitive higher education systems, elite universities perceive diversity as an advantage and compete with one another to bring in the best researchers and teaching professors, from both home and abroad. In less competitive systems, such as in many European nations, merit is not always the primary criteria for selection. The Italian context, which is characterized by high rates of favoritism in competitions for academic recruitment, is particularly suited for the investigation of gender differences in terms of possible discrimination and favoritism.

Our results show that, among candidates affected by negative bias, the incidence of female candidates is lower than that of male candidates. No gender differences occur among candidates who benefitted from positive bias. Our findings do not align with the prevalent opinions on the argument, as seen in the literature (e.g. Rotbart et al., 2012; Moss-Racusin et al., 2012; McGuire et al., 2004; Wright et al., 2003). There could be many reasons for this contrast. Without attempting to be exhaustive, we can suggest at least three: the use of different operative definitions of discrimination, different methods of measuring scientific merit, or the differing context of our analysis. However our results are in line with the expectations: if the concentration of male top scientists is greater than for women (Abramo et al., 2009b; Bordons et al., 2003), then one would in fact expect a greater concentration of discrimination against men. Furthermore, there is no evidence of gender differences among the favored winners, except in some disciplines where there is a greater concentration of male winners that would have not have merited the win compared to their female colleagues.

The results of the statistical analysis show that for the male candidates, the number of the candidate's career years in the same university as the committee members and the agreement between the gender of the applicant and that of the committee president assume greater weights in the judgment of competition outcomes than they do for the female applicants. However, the presence of a full professor in the same university with the same family name as the applicant assumes greater weight in the case of the female applicants than for the males.

The relation between academic candidates in the competitions analyzed and the total population of assistant professors in the SDSs involved is 12.8% for females (672/5,243) and 22.3% for males (1,642/7,373) and shows an unmistakable gender difference in the propensity to apply to the observed competitions, which could in part explain the phenomenon of the underrepresentation of women in the Italian academic systems, much more than any phenomena of bias as revealed by the current study.

**Appendix - Measuring research productivity**

Research activity is a production process in which the inputs consist of human resources and other tangible (scientific instruments, materials, etc.) and intangible (accumulated knowledge, social networks, etc.) resources, and where outputs have a complex character of both tangible nature (publications, patents, conference presentations, databases, protocols, etc.) and intangible nature (tacit knowledge, consulting activity, etc.). The new-knowledge production function thus has a multi-input and multi-output character. The principal efficiency indicator of any production system is labor productivity.

The calculation of labor productivity requires a few simplifications and assumptions. In the hard sciences, the prevalent form of codification of research output is publication in scientific journals. As a proxy of total output, in this work we consider only the specific publications (articles, article reviews, and proceeding papers) indexed in the Thomson Reuters WoS.

When measuring labor productivity, if there are differences in the production factors available to each scientist then one should normalize by them. Unfortunately relevant data are not available at the individual level in Italy. The first assumption then is that resources available to professors within the same field are the same. The second assumption is that the hours devoted to research are more or less the same for all professors. Given the characteristics of the Italian academic system as depicted in section 2.1, the above assumptions appear acceptable.

Because of the differences in the publication intensity across fields, a prerequisite of any distortion-free performance assessment is the classification of each researcher in one and only one field (Abramo et al. 2013a).

Most bibliometricians define productivity as the number of publications in the period of observation. Because publications have different values (impact), we prefer to adopt a more meaningful definition of productivity, i.e. the value of output per unit value of labor, all other production factors being equal. The latter recognizes that the publications embedding new knowledge have different value or impact on scientific advancement, which bibliometricians approximate with citations or journal impact factors. Provided that there is an adequate citation window (at least two years) the use of citations is always preferable (Abramo et al. 2011). Because citation behavior varies by field, we standardize the citations for each publication with respect to the average of the distribution of citations for all the Italian cited publications of the same year and the same WoS subject category[11]. Furthermore, research projects frequently involve a team of professors, which is registered in the co-authorship of publications. In this case we account for the fractional contributions of scientists to outputs, which is at times further signaled by the position of the authors in the byline.

At the individual level, professors of the same academic rank in this specific case, we can measure the average yearly productivity, named Fractional Scientific Strength (*FSS*), in the following way[12]:

$$FSS = \frac{1}{t} \sum_{i=1}^{N} \frac{c_i}{\bar{c}} f_i$$

Where:

$t$ = number of years of work of the researcher in the period of observation;

$N$ = number of publications of the researcher in the period of observation;

$c_i$ = citations received by publication $i$;

$\bar{c}$ = average of the distribution of citations received for all cited publications of the same year and subject category of publication $i$;

$f_i$ = fractional contribution of the researcher to publication $i$.

Fractional contribution equals the inverse of the number of authors in those fields where the practice is to place the authors in simple alphabetical order but assumes different weights in other cases. For the life sciences, widespread practice in Italy is for the authors to indicate the various contributions to the published research by the order of the names in the byline. For the life science SDSs, we give different weights to each co-author according to their order in the byline and the character of the co-authorship (intra-mural or extra-mural) (see Abramo et al. 2013b, 2013c). If first and last authors belong to the same university, 40% of citations are attributed to each of them; the remaining 20% are divided among all other authors. If the first two and last two authors belong to different universities, 30% of citations are attributed to first and last authors; 15% of citations are attributed to second and last author but one; the remaining 10% are divided among all others[13].

Data on research staff of each university, such as years of employment in the observed period, academic rank and their SDS classification are extracted from the database on Italian university personnel, maintained by the Ministry for Universities and Research[14]. Unfortunately, information on leaves of absence is not available and cannot be accounted for in the calculation of yearly productivity, to the disadvantage of women on maternity leave in the period of observation. The bibliometric dataset used to measure *FSS* is exctracted from the Italian Observatory of Public Research (ORP), a database developed and maintained by the authors and derived under license from the Thomson Reuters WoS. Beginning from the raw data of the WoS, and applying a complex algorithm for reconciliation of the author's affiliation and disambiguation of the true identity of the authors, each publication (article, article review and conference proceeding) is attributed to the university scientist or scientists that produced it (D'Angelo et al., 2011). Thanks to this algorithm we can produce rankings of research productivity at the individual level, on a national scale. Based on the value of *FSS* we obtain, for each SDS, a ranking list expressed on a percentile scale of 0-100 (worst to best) for comparison with the performance of all Italian colleagues of the same academic rank and SDS.

---

[13] The weighting values were assigned following advice from senior Italian professors in the life sciences. The values could be changed to suit different practices in other national contexts.

[14] http://cercauniversita.cineca.it/php5/docenti/cerca.php. Last access 26/10/2015.